\documentstyle[11pt,aaspp4,psfig]{article}

\def\kms{km~s$^{-1}$}
\def\HI{H\,{\sc i}}
\def\HII{H\,{\sc ii}}
\def\etal{{\rm et~al.\ }}

\begin{document}
\title{An unusual pulsar wind nebula associated with PSR~B0906--49}
\centerline{(To appear in {\em The Astrophysical Journal Letters})}
\author{B. M. Gaensler\altaffilmark{1,2}, B. W.
Stappers\altaffilmark{3, 4}, D. A. Frail\altaffilmark{5},
S. Johnston\altaffilmark{6}}
\altaffiltext{1}{Astrophysics Department, School of
Physics A29, University of Sydney, NSW 2006, Australia;
b.gaensler@physics.usyd.edu.au}
\altaffiltext{2}{Australia Telescope National Facility, CSIRO, PO Box 76,
Epping, NSW 2121, Australia}
\altaffiltext{3}{Astronomical Institute ``Anton Pannekoek'',
University of Amsterdam, Kruislaan 403, NL-1098 SJ Amsterdam, The Netherlands;
bws@astro.uva.nl}
\altaffiltext{4}{Mount Stromlo and Siding Spring Observatories, Institute
of Advanced Studies, Australian National University, Private Bag,
Weston Creek Post Office, ACT 2611, Australia}
\altaffiltext{5}{National Radio Astronomy Observatory, PO Box 0,
Socorro, NM 87801, United States of America; dfrail@nrao.edu.
NRAO is a facility of the National
Science Foundation operated under
cooperative agreement by Associated Universities, Inc.}
\altaffiltext{6}{Research Centre for Theoretical Astrophysics,
University of Sydney, NSW 2006, Australia; s.johnston@physics.usyd.edu.au}

\begin{abstract}

We report on Australia Telescope Compact Array observations of the
$\sim$$10^5$~yr old pulsar PSR~B0906--49. In an image containing only
off-pulse emission, we find a weak, slightly extended source coincident
with the pulsar's position, which we argue is best interpreted as a
pulsar wind nebula (PWN). A trail of emission extending behind the
pulsar aligns with the major axis of the PWN, and implies that
the pulsar is moving north-west with projected velocity
$\sim$60~\kms, consistent with its scintillation speed. The
consequent density we infer for the pulsar's environment is
$>2$~cm$^{-3}$, so that the PWN around PSR~B0906--49 is confined mainly
by the high density of its surroundings rather than by the pulsar's
velocity.  Other properties of the system such as the PWN's low
luminosity and apparent steep spectrum, and the pulsar's large characteristic
age, lead
us to suggest that this nebula is substantially different from other
radio PWNe, and may represent a transition between young pulsars with
prominent radio PWNe and older pulsars for which no radio PWN has been
detected.  We recommend that further searches for radio PWNe should be
made as here: at low frequencies and with the pulsed emission subtracted.

\end{abstract}


\section{Introduction}

The spin-down observed in most pulsars corresponds to a significant
rate of energy loss, which manifests itself primarily in the form of
a magnetized wind of relativistic particles
(e.g. Rees \& Gunn 1974\nocite{rg74}).  Under certain conditions
the interaction between this wind and its surroundings is observable,
in the form of a pulsar wind nebula (PWN).
At radio frequencies PWNe fall into two basic classes, plerions and
bow-shock nebulae: plerions (e.g.\ \cite{hb87})
are the filled-center components of supernova remnants (SNRs) in which
the pulsar wind is confined by the pressure of hot gas in the SNR
interior, while bow-shock nebulae are confined by the ram pressure
associated with their pulsar's high velocity (e.g.\ \cite{fk91};
\cite{fggd96}).  Both are characterized by significant levels of linear
polarization, a centrally-peaked morphology and a flat spectral index
($-0.3 < \alpha < 0$; $S_\nu \propto \nu^{\alpha}$).

While only a tiny fraction of a pulsar's spin-down energy goes into producing
the radio emission from such nebulae, radio PWNe are valuable diagnostics
of the properties of both the wind and of the pulsar itself
(e.g.\ \cite{fggd96}). Insight into the workings of such PWNe is
limited by the fact that only six pulsars, all with ages $\la10^5$~yr,
have associated radio PWNe (\cite{fs97}, hereafter FS97). Attempts to
detect radio nebulae around older pulsars have so far been unsuccessful (e.g.
\cite{ccgm83}). 

In an attempt to increase the sample of radio PWNe, FS97\nocite{fs97}
used the Very Large Array at 8.4~GHz to search for PWNe around 35
pulsars of high spin-down luminosity and/or velocity, but
found no new PWNe down to a surface
brightness $T_b \sim 1.2$~K.
They concluded that only young, energetic pulsars produce
observable radio nebulae, and that the properties of a pulsar's wind may
change as the pulsar ages.  However, 
a weak, compact or steep-spectrum nebula is difficult
to detect using the approach of FS97, because the nebula is ``hidden''
by the emission from its associated pulsar. 
This has motivated us
to attempt an alternative strategy for finding radio PWNe, whereby 
visibilities are recorded at high time resolution so that 
off-pulse images can be produced.
The full results of this program are described elsewhere (\cite{sgjf98});
here we report on the discovery of an unusual PWN associated with PSR~B0906--49.

PSR~B0906--49 ($l = 270\fdg3$, $b = -1\fdg0$) is a 107~ms pulsar
(\cite{dmd+88}) whose characteristic age $\tau_c =$~112\,000~yr and
spin-down luminosity $\dot{E} = 4.9 \times 10^{35}$~erg~s$^{-1}$ place
it amongst the 5\% youngest and most energetic of all pulsars.
\HI\ absorption has yielded distances to the pulsar in
the ranges 2.4--6.7~kpc (\cite{kjww95}) and 6.3--7.7~kpc
(\cite{sdw+96}), while a distance estimate using the pulsar's
dispersion measure (\cite{tc93}) is $6.6^{+1.3}_{-0.9}$~kpc. In future
discussion we assign it a distance $7d_7$~kpc. The pulse profile shows
a strong main pulse and weaker interpulse, both of
which are $\sim$90\% linearly polarized (\cite{wmlq93};
\cite{qmlg95}).  Scintillation measurements imply a transverse velocity
for the pulsar of $\sim$50$d_7^{1/2}$~\kms\ (\cite{jnk98}).

\section{Observations}

Observations of PSR~B0906--49 were carried out with the Australia
Telescope Compact Array (ATCA), 
and are described in Table~\ref{tab_observations}.
In each observation, two continuum frequencies were observed
simultaneously with 32 4-MHz channels across each; all four Stokes
parameters were recorded. All observations were carried out in pulsar
gating mode, in which visibilities were sampled at a time resolution
of 3.3~ms (32 bins per period) 
and then folded at the apparent pulse period.
These summed visibilities were then recorded every 40 seconds.

The flux density scale of the observations was determined by
observations of PKS~B1934--638, while polarization and antenna gains 
were calibrated using observations every 45~min of
PKS~B0823--500. Data were edited and calibrated using the MIRIAD
package, and then dedispersed at a dispersion measure
of 179~pc~cm$^{-3}$. The resultant smearing
due to the finite channel width was in all cases no more than
one time bin. At each frequency the field of
interest was imaged, deconvolved using $10^4$ iterations of the CLEAN
algorithm, and then convolved with a Gaussian beam of
dimensions corresponding to the diffraction limit.

\section{Results}

\begin{figure}
\centerline{\psfig{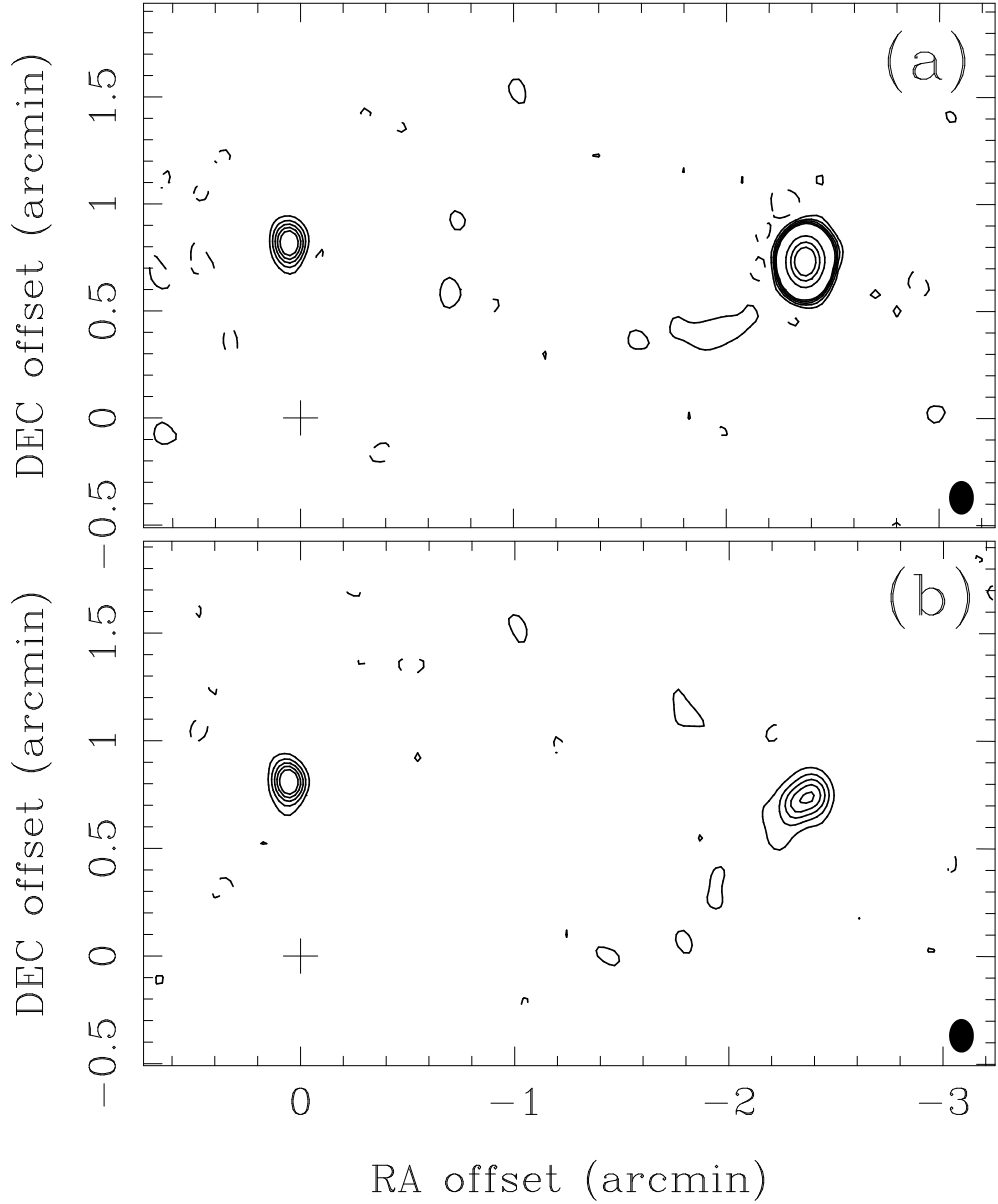}}
\caption{1.3-GHz images (1997 May and 1997 Aug data combined)
of a region surrounding PSR~B0906--49,
with baselines shorter than 1~k$\lambda$ excluded:  (a)
all time bins; (b) off-pulse bins only. The rms
noise in the images is limited by confusion to
0.06~mJy~beam$^{-1}$ ($T_b = 0.7$~K). Contours
are at --0.15 (broken), 0.15, 0.3, 0.45, 0.6, 0.75, 5, 10 and
15~mJy~beam$^{-1}$. The resolution (as shown at the
bottom right of each image) is $9\farcs4  \times
7\farcs0$, PA 0\arcdeg. The
``+'' indicates the position of the phase center,  RA (J2000) $09^{\rm
h}08^{\rm m}50^{\rm s}$, Dec (J2000) $-49\arcdeg13\arcmin50\arcsec$.}
\label{fig_contours_1344}
\end{figure}

\begin{figure}
\centerline{\psfig{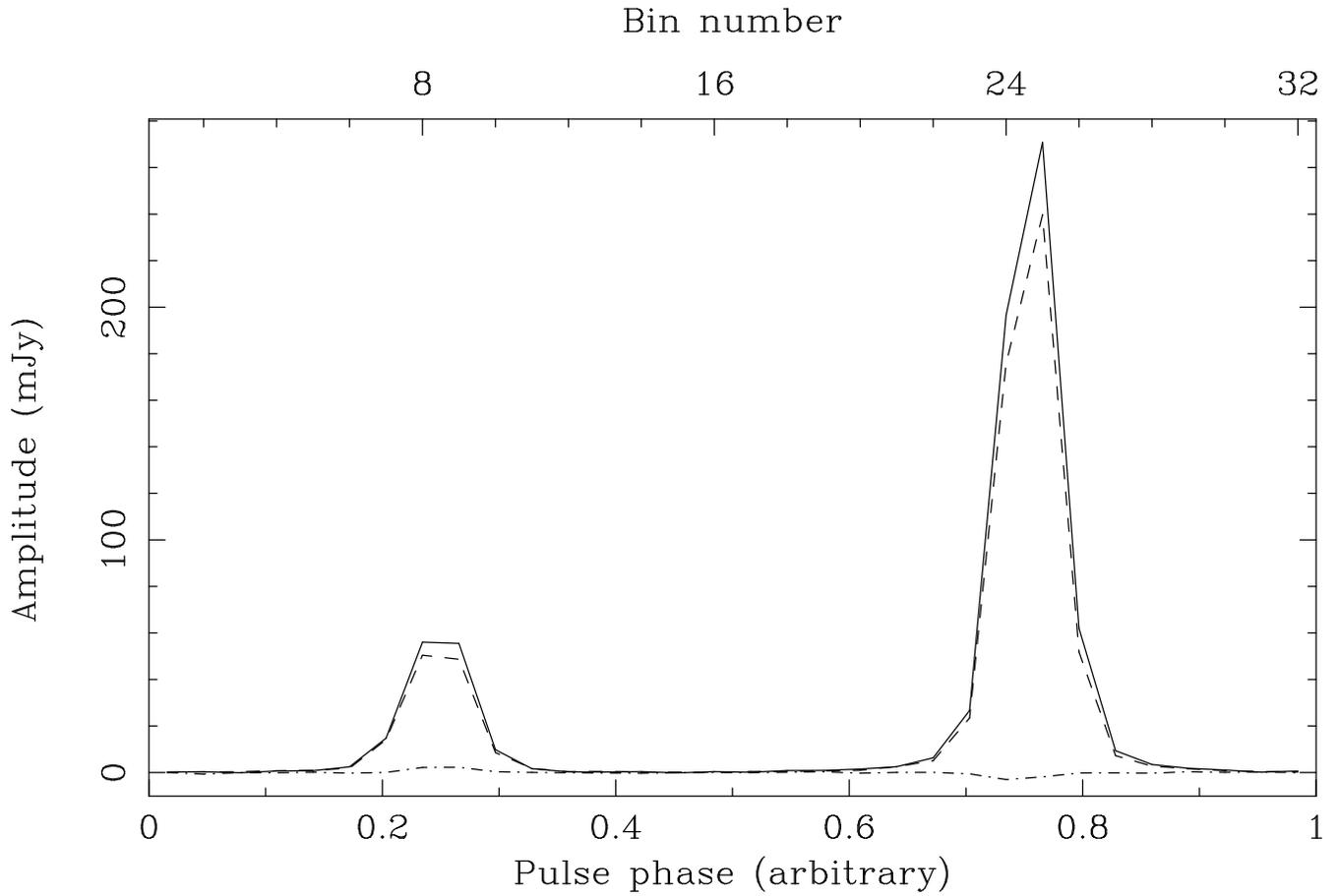}}
\caption{Pulse profile of B0906--49 at 1.3~GHz. The solid,
broken and dot-dashed lines represent total, linearly polarized
and circularly polarized intensity respectively.
The profile is 90\% linearly and 3\% circularly polarized,
values close to those determined in previous measurements.}
\label{fig_profile}
\end{figure}

In Fig~\ref{fig_contours_1344}(a) we show a 1.3-GHz image of the region
surrounding the pulsar, using all time bins from 1997 May and
1997 Aug observations.
An adjacent \HII\ region (\cite{gcly98}) 
contributes considerable extended emission, which we
discard by imaging using only $u-v$ spacings greater than
1~k$\lambda$.  The pulsar is apparent as an unresolved source at
position  RA (J2000) $09^{\rm h}08^{\rm m}35^{\rm s}$, Dec (J2000)
$-49\arcdeg13\arcmin06\arcsec$, in agreement with its timing
position (\cite{dmd+88}).  Extracting the time dependence
of the data at these coordinates from the visibilities gives the
profile shown in Fig~\ref{fig_profile}, clearly showing the main
pulse and interpulse. Similar profiles are obtained
at other frequencies: flux densities for the pulsar (including any
coincident unpulsed emission) are given in Table~\ref{tab_fluxes}.
An unrelated source, ATCA~J090850--491301,
can be seen 2\farcm5 to the east of the pulsar, 
with flux densities also listed in Table~\ref{tab_fluxes}.

Fig~\ref{fig_contours_1344}(b) shows the same field as in
Fig~\ref{fig_contours_1344}(a), but generated by only using time bins in which there
is no discernible pulsed emission. A weak source can now be seen at the
position of the pulsar. While the pulsar's emission is well-fitted by a
Gaussian corresponding to a point source, a fit to this weaker source
shows it to be extended, with dimensions $15\farcs0 \times 10\farcs0$
at position angle 315\arcdeg$\pm$10\arcdeg\ (N through E).  After
correction for bandwidth depolarization, the source
is less than $\sim$50\% polarized. The position of its peak agrees
with that of PSR~B0906--49 to 0\farcs6 in RA and to 0\farcs4 in Dec.

The flux density of the off-pulse source as a function of frequency is
detailed in Table~\ref{tab_fluxes}.  In observations at 1.7~GHz and
2.2~GHz the source was not detected, despite being simultaneously
detected at each epoch at a lower frequency. For non-detections we
quote upper limits to the flux density, derived using  the method of
Green \& Scheuer (1992\nocite{gs92}). This involves adding to the data
simulated sources of the same dimensions as the off-pulse source,
increasing the flux density until the simulated source is
distinguishable above the noise. The upper limits quoted in
Table~\ref{tab_fluxes} are conservative, being twice the values at
which a simulated source could be discerned.

\begin{figure}
\centerline{\psfig{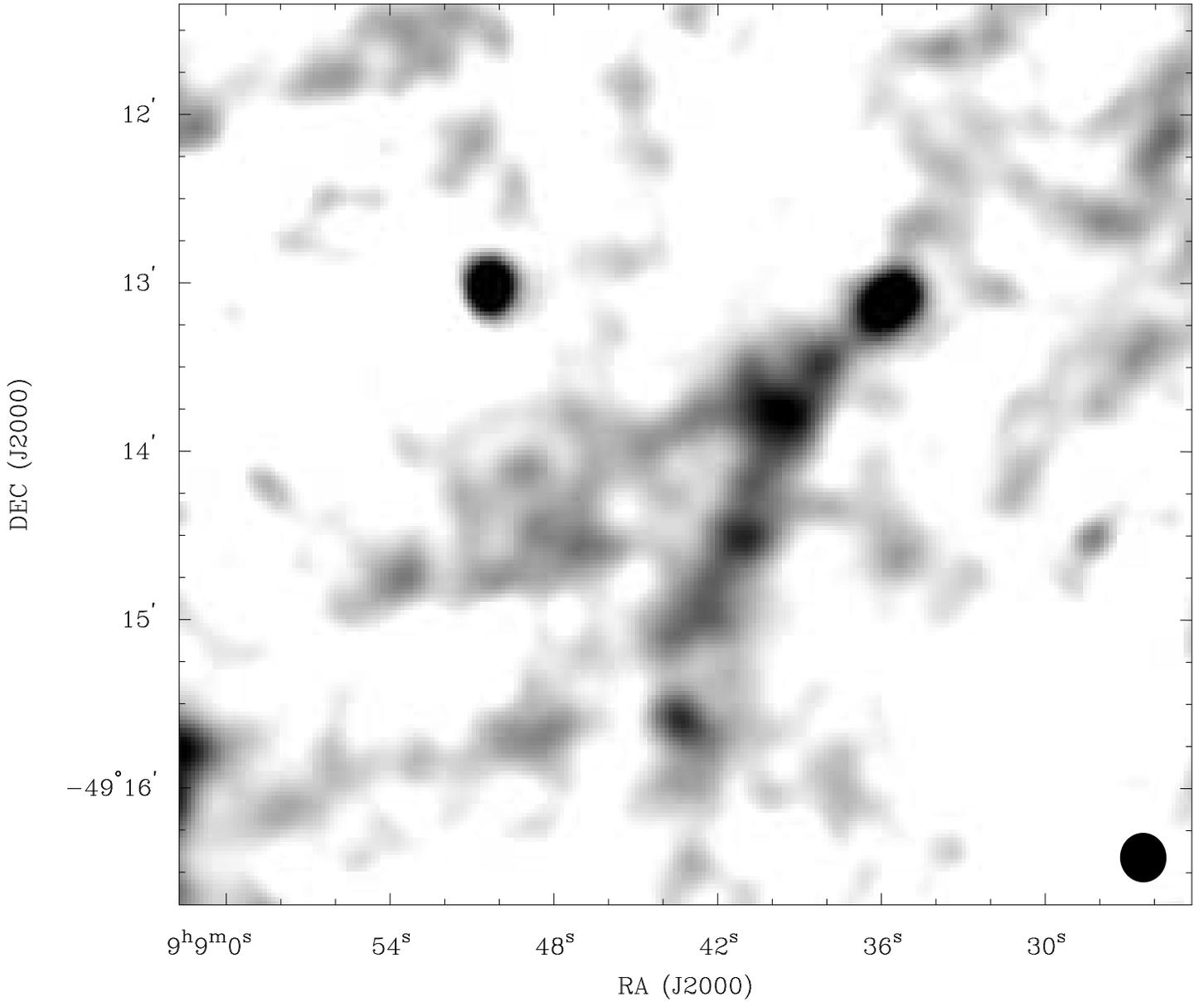}}
\caption{A 1.3-GHz image of off-pulse emission from 1997 May 12 data,
using all baselines.
The gray-scale range is 0 to 0.5
mJy~beam$^{-1}$, and the resolution (indicated at bottom right) is
$17\farcs7 \times 16\farcs6$, PA 5\arcdeg.
The rms noise is 0.12~mJy~beam$^{-1}$.}
\label{fig_trail}
\end{figure}

A 1.3-GHz off-pulse image from 1997 May 12 data, made using all
baselines, is shown in Fig~\ref{fig_trail}.  As well as
ATCA~J090850--491301 and the off-pulse source, a faint trail of
emission can be seen, extending $\sim$$3\farcm5$ south-east from the
pulsar's position, broadening with increasing distance from it before
abruptly terminating. The apex of the trail coincides with the position
of the pulsar, and the position angle defined by the bisection of the
trail's two flanks is consistent with the direction along which the
off-pulse source is extended. The flux density of the trail is
$\sim$7~mJy.  In 1.3-GHz data from 1997 Aug 17 observations, the trail
is just visible amidst confusing thermal emission. The trail was not
detected in 1.2-GHz observations  (which were severely affected by
interference), nor was it seen at 1.7 or 2.2~GHz (as for the off-pulse
source).  While the inclusion of short baselines produces a great deal of
confusing structure, no other feature resembling the trail is
seen in any part of the field in any observation.

An H$\alpha$ observation of the region (\cite{bbw98}) shows no emission
which can be associated with any radio-emitting source, a result which
can be attributed to a foreground dark cloud (\cite{hms+86}).

\section{Discussion}

We consider four possible explanations for the weak source coincident
with PSR~B0906--49: 
positional coincidence with an unrelated source, 
a compact SNR, 
emission from the pulsar itself, 
or a pulsar wind nebula associated with the pulsar.

At $\lambda=20$~cm, extragalactic source counts (e.g. Windhorst et al.
1985\nocite{wmo+85}) imply a probabibility $\sim$$10^{-5}$ of finding a
background radio source of flux density $>$1~mJy within 1\arcsec\ of
the pulsar, allowing us to rule out a coincidental alignment. We
estimate that positional coincidence with an unrelated Galactic source
is also unlikely.

The source under consideration may be a compact SNR, in the form of
either a shell (\cite{gre85}) or a plerion (\cite{hvbl89}).  However, a
shell or plerion of age $\tau_c \approx 10^5$~yr and radius $\approx
0.3 d_7$~pc as required here results in an improbably high
($>10^8$~cm$^{-3}$) density for the surrounding medium.
No other emission is seen in the region which might correspond to a
more extended SNR (\cite{dshj95}; \cite{gcly98}), and we presume it to
have dissipated.

Although we took care to create an image using only off-pulse bins, it
is impossible to exclude the possibility that
Fig~\ref{fig_contours_1344}(b) still contains some flux from the
pulsar, in the form of either unpulsed emission or a further weak 
component to the pulse profile.  Our main argument against such interpretations 
is that the off-pulse source appears extended.  Although the source is too
weak to fit to in the $u-v$ plane, its extension and position angle are
quite different from that of the synthesized beam, and are visible at
all reasonable contour levels.  It is difficult to see how the presence
of noise, sidelobes or other artifacts could artificially create such
an appearance. The extension is not directed towards the phase center
and so cannot be accounted for by bandwidth smearing. Nor is there
evidence for significant time variability in the pulsar's flux, which might
also mimic an extended source (and in any case would cause the 
pulsar itself to appear similarly extended).
The source is evident in repeated observations, and
persists when imaging various subsets of the off-pulse component of the
pulse profile.  The off-pulse source also lacks the high fractional
linear polarization observed for the pulsar. We thus conclude that it
is unlikely that the source represents emission from the pulsar
itself.

The remaining possibility is that this source
represents synchrotron emission from a pulsar wind nebula
associated with PSR~B0906--49.
The observed elongation of the PWN suggests that it is a bow-shock
nebula, an interpretation supported by the trail's morphology and its
alignment with the PWN's major axis.  We interpret the trail as a
synchrotron wake\footnote{In further discussion the ``PWN'' corresponds
to the compact off-pulse source, while the ``trail'' corresponds to the
extended emission to the south-east.}
extending behind the pulsar, whose inferred direction
of motion is thus along PA 315\arcdeg.
This wake is analogous to
similar radio structures seen trailing behind PSRs B1757--24
(\cite{fk91}) and B1853+01 (\cite{fggd96}) and behind the X-ray binary
Circinus X-1 (\cite{schn93}).

By demanding that the pulsar has traversed the trail's length in its
lifetime (which we equate to $\tau_c$ in all further discussion), we
determine a velocity $v \ge 60 d_7 / \sin i $~\kms, where $i$ is the
inclination of the pulsar's direction of motion to the line of sight.
The well-defined edge to the trail in the south-east suggests that the
lack of emission beyond this point corresponds to a change in
conditions rather than to synchrotron losses (which are most likely
negligible over the pulsar's lifetime in any case). Given the agreement
between the lower limit on $v$ and the scintillation velocity (Johnston
et al. 1998\nocite{jnk98}), we propose that the pulsar was born at the
base of the trail and adopt a velocity $v = 60v_{60}/ \sin i$~\kms.
The implied proper motion of $1.8 v_{60} d_7^{-1}$~mas~yr$^{-1}$ is too
small to be measured directly, but could
be constrained or ruled out through pulsar timing or VLBI.

The pulsar wind flows freely from the pulsar until it encounters a
reverse shock at radius~$r_s$ (e.g.\ \cite{cor96}). 
The location of this shock is unresolved in our data, and from
our highest resolution data we
estimate $r_s < 10^{17} d_7$~cm.  We assume that all the
pulsar's spin-down luminosity goes into the pulsar wind; upper limits
to its $\gamma$-ray flux (\cite{tab+94}) make this a reasonable
assumption to within a factor of 2.  Thus, equating the energy density
of the pulsar wind, $\dot{E}/4\pi r_s^2 c$, with the ram pressure due
to the pulsar's motion, $\rho v^2$, we find an ambient (hydrogen) density $n >
2(d_7 v_{60}/\sin i)^{-2}$~cm$^{-3}$, significantly higher than for the
general interstellar medium.  A distance greater than 7~kpc is
inconsistent with the pulsar's \HI\ absorption; thus only if the
pulsar's velocity is significantly greater than 60~\kms\
can this discrepancy be reconciled. If we adopt the lower distance limit
determined by Koribalski \etal\ (1995\nocite{kjww95}) and adjust
the pulsar velocity accordingly, the required density rises to
$n > 70/ \sin i$~cm$^{-3}$.  We note that the CO survey of Grabelsky
\etal\ (1987\nocite{gcb+87}) shows molecular material in the
direction of PSR~B0906--49, at velocities corresponding to the range
determined for the pulsar from \HI\ absorption.  Thus PWN~B0906--49
appears to be a bow-shock PWN in which the
main contribution to the confining ram pressure appears to be the
density of surrounding material rather than the pulsar's high velocity.
While there are other instances of bow-shock PWNe around low velocity
pulsars (see \cite{cor96}), PSR~B0906--49 is the only case in which
the PWN has been detected at radio frequencies.

The spectral index of PWN~B0906--49 is poorly determined by the available
data, but assuming a single power law over the observed
range implies $-3.5 \la \alpha \la -0.7$. This is significantly steeper than
for filled-center SNRs (\cite{gre96b}) and for the six other known
radio PWN; we note that such a nebula would not have been detected by
the 8.4~GHz search of FS97.  For $\alpha = -1$, the integrated
radio luminosity of the PWN is $L_R \approx 10^{30}
d_7^2$~erg~s$^{-1}$, corresponding to an efficiency $\epsilon_R \equiv L_R /
\dot{E} \approx 2 \times 10^{-6} d_7^2$.  These values of 
$L_R$ and $\epsilon_R$ 
are significantly lower than for any
other radio PWN (see FS97\nocite{fs97}). Of pulsars powering such
nebulae, only PSR~B1853+01 has a (slightly) lower $\dot{E}$. 

The unusual spectrum of PWN~B0906--49 may relate to a decrease in the
efficiency of shock acceleration resulting from enhanced densities and
reduced shock velocities (\cite{plsr97}).  If a significant component
of the radio emission is due to electrons accelerated at the shock
front between the pulsar wind and its surroundings, the high density
and low velocity we have proposed for this source might explain its
steep spectrum and low luminosity. Alternatively, FS97 have argued that
older, less energetic pulsars do not produce observable radio nebulae,
and a simple explanation for both their and our observations is that
the intrinsic spectrum of injected electrons steepens with age.  Apart
from PSR~B1951+32 ($\tau_c = 107\,000$~yr), B0906--49 is at least five
times older than any other pulsar associated with a radio PWN. While
PSR~B1951+32 is only slightly younger than PSR~B0906--49, the former
system represents a complicated interaction between the pulsar and its
associated SNR (\cite{sfs89}), and it is difficult to separate the
contribution of the pulsar wind from that of the supernova shock.

\section{Conclusion}

Using pulsar-gating observations at $\lambda=20$~cm, we have found an
off-pulse, slightly extended source coincident with PSR~B0906--49,
which we interpet as a faint pulsar wind nebula. An associated trail
implies a projected velocity for the pulsar of $\sim$60~\kms\ along a position
angle 315\arcdeg.  The system is unusual in several ways:

\begin{enumerate}

\item the nebula has a lower luminosity and steeper
spectrum than any other radio PWN yet discovered;

\item PSR~B0906--49 is older than any other pulsar known to power a
radio PWN, while only PSR~B1853+01 has a lower spin-down
luminosity;

\item the PWN appears to be generated by a slow ($\sim$60~\kms)
pulsar moving through a dense ($>2$~cm$^{-3}$) medium.
\end{enumerate}

Thus PWN~B0906--49 appears to be very different from other radio
PWNe. To account for this, we
propose that either the low velocity and high density inhibits shock
acceleration, or that the particle spectrum injected by a pulsar into
its PWN steepens with age. In the latter case, PWN~B0906--49 may be in
transition between the high luminosity, flat-spectrum PWNe seen
around young pulsars, and the undetectable radio PWNe around older pulsars.
We note that PWNe of low flux density and steep
spectrum were heavily selected against in FS97's ungated, high
frequency observations. We therefore recommend that future searches,
particularly around intermediate-age pulsars, should be made as here:
at low frequencies, and by imaging only the off-pulse emission.  A
collection of PWNe produced under a variety of conditions still needs
to be accumulated before we can fully understand how the radio emission
from a PWN is produced and how it depends on the properties of its
pulsar. 

\acknowledgements

We thank the ATNF Time Assignment Committee for generously assigning an
extra configuration, Mike Bessell and Michelle Buxton for providing
H$\alpha$ data on the region, and Dick Manchester for supplying
the pulsar ephemeris.  BMG acknowledges the support of an
Australian Postgraduate Award, while BWS received support from an ANU
PhD Scholarship and the ATNF Higher Degree Programme.  The Australia
Telescope is funded by the Commonwealth of Australia for operation as a
National Facility operated by CSIRO.  This research has made use of
NASA's Astrophysics Data System Abstract Service and of the SIMBAD
database, operated at CDS, Strasbourg, France.

\clearpage


\clearpage

\begin{table}
\begin{tabular}{cccccc} \tableline \tableline
Date        & Array         & Maximum      & Time On &
\multicolumn{2}{c}{Center Frequency (GHz)} \\
            & Configuration & Baseline (m) & Source (h) & 1 & 2 \\ \tableline
1997 May 12 &  6.0B & 5969 &  12 & 1.344  & 2.240 \\
1997 Aug 17 &  1.5B & 1286 &  12 & 1.344  & 2.240 \\
1997 Nov 06 &  6.0C & 6000 &  11 & 1.216  & 1.728 \\ \tableline
\end{tabular}
\caption{ATCA observations of PSR~B0906--49.}
\label{tab_observations}
\end{table}

\begin{table}
\begin{tabular}{ccccc} \tableline \tableline
Frequency        &  \multicolumn{3}{c}{Flux Density (mJy)\tablenotemark{a}} &
Rms Noise \\
(GHz)		 &  PSR~B0906--49   & PWN~B0906--49  & ATCA~J090850--491301 &
(mJy~beam$^{-1}$)\tablenotemark{a} \\
\tableline
1.216            &      22$\pm$2      &  1.4$\pm$0.3      &   1.0$\pm$0.2 &
0.14 \\
1.344            &      25$\pm$2      &  1.5$\pm$0.3  & 1.1$\pm$0.2 &
0.06 \\
1.728            &      16$\pm$2      & $<1.0$ & 1.0$\pm$0.2  & 0.07 \\
2.240            &      14$\pm$2      & $<1.0$ & 1.0$\pm$0.2 & 0.06 \\ \tableline
\end{tabular}
\tablenotetext{a}{1~Jy $= 10^{-26}$ W m$^{-2}$ Hz$^{-1}$} 
\caption{Flux densities and sensitivity as a function of frequency.
On the basis of simultaneous dual frequency
observations, we compute an approximate spectral index for the pulsar
of $\alpha \sim -1$. The differences in the pulsar's flux density 
between the three
observations are most likely due to refractive scintillation.}

\label{tab_fluxes}
\end{table}

\clearpage

\clearpage

\clearpage

\end{document}